\documentclass[letterpaper, 10 pt, conference]{ieeeconf}  

\IEEEoverridecommandlockouts                              
\usepackage{graphics} 
\usepackage{epsfig} 
\usepackage{mathptmx} 
\usepackage{times} 
\usepackage{amsmath} 
\usepackage{amssymb}  
\usepackage{tcolorbox}
\usepackage{pgfkeys}
\usepackage{xcolor}
\usepackage{pgfplots}
\usepackage{layouts}
\usepackage{tikz}
\usepackage[ngerman]{babel}
\usepackage{algorithm}
\usepackage[noend]{algpseudocode}
\usepackage{mathtools}

\makeatletter
\newcommand\fs@spaceruled{\def\@fs@cfont{\bfseries}\let\@fs@capt\floatc@ruled
	\def\@fs@pre{\vspace{0.55\baselineskip}\hrule height.8pt depth0pt \kern2pt}%
	\def\@fs@post{\kern2pt\hrule\relax}%
	\def\@fs@mid{\kern2pt\hrule\kern2pt}%
	\let\@fs@iftopcapt\iftrue}
\makeatother

\newcommand\copyrighttext{%
	\footnotesize \copyright 2024 IEEE. Personal use of this material is permitted. Permission from IEEE must be obtained for all other uses, in any current or future media, including reprinting/republishing this material for advertising or promotional purposes, creating new collective works, for resale or redistribution to servers or lists, or reuse of any copyrighted component of this work in other works.}
\newcommand\copyrightnotice{%
	\begin{tikzpicture}[remember picture,overlay]
		\node[anchor=south,yshift=7pt] at (current page.south) {\fbox{\parbox{\dimexpr\textwidth-\fboxsep-\fboxrule\relax}{\copyrighttext}}};
	\end{tikzpicture}%
}

\shorthandon{"}
\def \iIOSS/{i"~IOSS}
\shorthandoff{"}

\definecolor{new}{rgb}{0,0,0}
\newlength\figurewidth

\title{\LARGE \bf
Event-triggered moving horizon estimation for nonlinear systems
}

\author{Isabelle Krauss, Julian D. Schiller, Victor G. Lopez, and Matthias A. Müller
	\thanks{This work received funding from the European Research Council (ERC) under the European Union’s Horizon 2020 research and innovation programme (grant agreement No 948679).}
	\thanks{I. Krauss,  J. D. Schiller, V. G. Lopez and M. A. Müller are with the Leibniz University Hannover, Institute of Automatic Control,  30167 Hannover,  Germany
		{\tt\small \{krauss,schiller,lopez,mueller\}\newline@irt.uni-hannover.de}}
}

\begin{document}
		\newtheorem{thm}{Theorem}
\newtheorem{cor}{Corollary}
\newtheorem{lem}{Lemma}
\newtheorem{prop}{Proposition}

\newtheorem{rem}{Remark}

\newtheorem{defi}{Definition}
\newtheorem{ass}{Assumption}

\setlength\figurewidth{0.9\columnwidth} 

\maketitle
\thispagestyle{empty}
\pagestyle{empty}
\copyrightnotice

\begin{abstract}
This work proposes an event-triggered moving horizon estimation (ET-MHE) scheme for general nonlinear systems. 
The key components of the proposed scheme are a novel event-triggering mechanism (ETM) and the suitable design of the MHE cost function.
The main characteristic of our method is that the MHE's nonlinear optimization problem is only solved when the ETM triggers the transmission of measured data to the remote state estimator.
If no event occurs, then the current state estimate results from an open-loop prediction using the system dynamics. Furthermore, we show robust global exponential stability of the ET-MHE  under a suitable detectability condition.
Finally, we illustrate the applicability of the proposed method in terms of a nonlinear benchmark example, where we achieved similar estimation performance compared to standard MHE using 86\%  less computational resources.
\end{abstract}

\section{INTRODUCTION}
\label{sec:Intro}
 Moving horizon estimation (MHE) is an optimization-based state estimation technique. Its central feature is the minimization of a cost function while taking past measurements into account to obtain an estimate of the current state.  In recent years, MHE has gained increasing attention due to its applicability to uncertain nonlinear and potentially constrained systems subject to model inaccuracies and measurement noise. Robust stability guarantees for MHE have been established under a mild detectability condition (incremental input/output-to-state stability (\iIOSS/)), cf. \cite{All21,Ji16,Knu23,Mul17,Sch23}.  In \cite{Sch23} a Lyapunov characterization of \iIOSS/ was utilized to show robust stability. Furthermore, a method  was developed to verify \iIOSS/ for classes of nonlinear systems by computing an \iIOSS/ Lyapunov function. 
In the context of state estimation, limited resources such as computation power, energy, and communication bandwidth can pose a challenge in  applications such as networked control systems.
Such systems are composed of a large number of devices (sensors, actuators, etc.) interconnected by communication networks. 
Frequent signal  transmission can be a challenge for wireless communication. Due to limited bandwidth, not all components  may access the communication channel and transmit data simultaneously \cite{Zou21}. Such limitations can be addressed by event-triggering strategies. In addition, components such as sensors and communication devices powered by batteries with limited capacity encourage the use of event-triggering mechanisms (ETM) to reduce the power consumption \cite{Aky02}.
\par For linear systems there are some notable results on event-triggered optimization-based state estimation. An event-triggered maximum likelihood state estimation method for detectable systems in the presence of Gaussian noise was developed  in \cite{Shi14}. In \cite{Yin20}, \cite{Zou20} and \cite{LiXu23} the authors proposed event-triggered moving horizon estimation (ET-MHE) schemes. In all three cases, boundedness of the estimation error was shown for observable systems and bounded disturbances.  
\par Several event-triggered state estimation methods capable of handling nonlinear systems rely on nonlinear extensions of the Kalman filter. For instance,  a cubature Kalman filter with a deterministic event-triggered scheduling strategy was presented in \cite{Li21}. An unscented and a cubature Kalman filter using stochastic event-trigger conditions were proposed in \cite{Li19} and \cite{Li23}, respectively. There, stochastic stability was shown under some observability assumption. Apart from such Kalman filter-based approaches, in \cite{Zou21} a nonlinear MHE  was developed for networked  systems with  linear output functions and sector-bounded nonlinear system dynamics using a so-called random access protocol to schedule data transmission.
\par 
This paper proposes an ET-MHE scheme for remote state estimation of general nonlinear detectable systems.
We develop an event-triggering condition involving a design variable to tune its sensitivity.
This reduces the frequency at which (i) the optimization problem is solved, and (ii) a communication channel  connecting the system with a remote estimator needs to be accessed for data transmission. Then, we establish robust global exponential stability (RGES) of the estimation error with respect to disturbances and noise.  Here, the minimum required horizon length does not depend on the event-triggering condition but is the same as for MHE without event-triggering.  
\par The rest of the paper is organized as follows. In Section \ref{sec:Setup}  we explain the setting of this work and provide some technical definitions.
Section \ref{sec:MHEScheme} presents the ET-MHE scheme, in particular the ETM and the MHE objective. Robust stability of the proposed ET-MHE under a suitable detectability condition is then shown in Section \ref{sec:StabAna}. 
This is followed by a simulation example illustrating the effectiveness of the event-triggered MHE scheme in Section \ref{sec:NumEx}.
This is followed by a simulation example  in Section 3 illustrating, as previously indicated, the effectiveness of the event-triggered MHE scheme.
\subsubsection*{Notation} 
The set of all nonnegative real numbers is denoted by $\mathbb{R}_{\geq0}$. We denote the set of integers in the interval $[a,b]$  for some  \mbox{$a,b \in \mathbb{R}$}  by $\mathbb{I}_{[a,b]}$ and the set of integers greater or equal to $a$  for some  $a\in \mathbb{R}$  by $\mathbb{I}_{\geq a}$.  The bold symbol $\mathbf{u}$ refers to a sequence of the vector-valued variable \mbox{${u\in\mathbb{R}^m}, \ \mathbf{u}= \{u_0,u_1,\ldots\}$}.
The Euclidean norm of vector $x \in \mathbb{R}^n$  is denoted by $||x||$ and $||x||_P^2=x^\top Px$ for a positive definite matrix $P=P^\top$. Let $\lambda_{\text{min}}(P)$ and  $\lambda_{\text{max}}(P)$ be the minimum and maximum eigenvalue of $P$, respectively. The maximum generalized eigenvalue of positive definite matrices $P,Q$ is represented by $\lambda_{\text{max}}(P,Q)$. $P\succ 0$ ($P\succeq 0$) denotes a positive definite (positive semi-definite) matrix.
\section{Preliminaries and Problem Setup}
\label{sec:Setup}
\par We consider the discrete-time nonlinear system 
\begin{align}
	\begin{aligned}
		x_{t+1}&=f(x_t,u_t,w_t) \\
		y_t&=h(x_t,u_t,w_t)\\
	\end{aligned}
	\label{eq:sys}
\end{align}
with state $x_t \in \mathbb{X} \subseteq \mathbb{R}^n$, control input $u_t\in \mathbb{U} \subseteq \mathbb{R}^m$, process disturbance and measurement noise  $w_t \in \mathbb{W} \subseteq \mathbb{R}^q$ with \mbox{$0 \in \mathbb{W}$}, noisy output measurement $y_t \in \mathbb{Y} \subseteq \mathbb{R}^p$, time $t \in \mathbb{I}_{\geq 0}$, and nonlinear continuous functions $f: \mathbb{X} \times \mathbb{U} \times \mathbb{W} \rightarrow \mathbb{X}, \ h: \mathbb{X} \times \mathbb{U} \times \mathbb{W}\rightarrow \mathbb{Y}$ representing the system dynamics and the output model, respectively.
\par In order to design a robustly stable MHE, some suitable detectability assumption of the system is required. \textcolor{new}{In recent years, \iIOSS/ has become standard  as a description of nonlinear detectability in the context of MHE.
Since we aim to establish robust global exponential stability, we use an exponential version of  \iIOSS/ as a detectability condition.}
\begin{ass}[Exponential \iIOSS/]
	The system (\ref{eq:sys})  is exponentially \iIOSS/, i.e., there exist $P_1,P_2 \succ 0$, $Q, R \succeq 0$ and $\eta \in [0, 1)$ such that for any input trajectories  $\mathbf{u}\in\mathbb{U}^\infty$ and $\mathbf{w},\mathbf{\tilde{w}}\in\mathbb{W}^\infty$ and any pair of initial conditions $x_{0}$, $\tilde{x}_{0}\in \mathbb{X}$ it holds for all $t\geq 0$
	\begin{align}
		\begin{aligned}
			||x_t-\tilde{x}_t||_{P_1}^2&\leq ||x_0-\tilde{x}_0||_{P_2}^2 \eta^t +\sum_{j=0}^{t-1}\eta^{t-j-1}||w_j-\tilde{w}_j||_Q^2\\
			&+\sum_{j=0}^{t-1}\eta^{t-j-1}||y_j-\tilde{y}_j||_R^2,
		\end{aligned}
		\label{eq:ioss}
	\end{align}
	\label{ass:eIOSSET}
	where $x_{t+1}=f(x_{t},u_{t},w_{t})$ and $\tilde{x}_{t+1}=f(\tilde{x}_{t},u_{t},\tilde{w}_{t})$ as well as $y_t=h(x_{t},u_{t},w_{t})$ and $\tilde{y}_t=h(\tilde{x}_{t},u_{t},\tilde{w}_{t})$ for all $t\geq 0$.
\end{ass}
\par Note that Assumption \ref{ass:eIOSSET} is necessary for the existence of robustly exponentially stable state estimators (cf. \cite[Prop. 3]{Knu20}, \cite[Prop. 2.6]{All20}).
 By adapting the converse Lyapunov theorem from \cite{All20}, Assumption \ref{ass:eIOSSET} is equivalent to the system (\ref{eq:sys}) admitting a quadratically-bounded \iIOSS/ Lyapunov function $W_{\delta}(x,\tilde{x})$ that satisfies for all $x,\tilde{x} \in \mathbb{X}$, all $u\in \mathbb{U}$, all $w,\tilde{w} \in \mathbb{W}$ and all $y,\tilde{y} \in \mathbb{Y}$ with $y=h(x,u,w)$ and $\tilde{y}=h(\tilde{x},u,\tilde{w})$
\begin{align}
	\begin{aligned}
&	||x-\tilde{x}||_{P_1}^2\leq W_{\delta}(x,\tilde{x})\leq	||x-\tilde{x}||_{P_2}^2,\\
&	W_{\delta}(f(x,u,w),f(\tilde{x},u,\tilde{w}))\\ & \quad \leq \eta W_{\delta}(x,\tilde{x})+ 	||w-\tilde{w}||_{Q}^2+	||y-\tilde{y}||_{R}^2 
\end{aligned}
\label{eq:Lyap}
\end{align}	
with $\eta \in [0, 1)$, $P_1,P_2 \succ 0$ and $Q,R \succeq 0$.
A systematic method to compute an  \iIOSS/ Lyapunov function and thus to verify Assumption~\ref{ass:eIOSSET} is proposed in \cite{Sch23}.
\par In Section \ref{sec:StabAna}, we establish the following stability property of the proposed ET-MHE.
\begin{defi}[RGES {\cite[Def. 1]{Knu18}}]
	A state estimator for system (\ref{eq:sys}) is robustly globally exponentially stable (RGES) if there exist $C_x,C_w>0$ and $\lambda_x,\lambda_w \in [0,1)$  such that for any initial conditions $x_0,\hat{x}_0 \in \mathbb{X}$ and any disturbance sequence~$\mathbf{w}\in\mathbb{W}^\infty$ the resulting state estimate $\hat{x}_t$ satisfies the following  for all $t\geq 0$
	\begin{align}
		\begin{aligned}
			||x_t-\hat{x}_t||\leq C_x||x_0-\hat{x}_0||\lambda_x^t+\sum_{j=0}^{t-1}
			C_w||w_j||\lambda_w^{t-j-1}.
		\end{aligned}
		\label{eq:rges}
	\end{align} 
	\label{def:rges}
\end{defi}
\par Note that in this definition of RGES, the influence of the past disturbances on the error bound is discounted. Hence, it directly implies that the estimation error converges to zero for vanishing disturbances. 
We point out that Definition~\ref{def:rges} is equivalent to the max-based formulation of RGES (that is, (\ref{eq:rges}) with each sum operation replaced by maximization), cf. \cite[Proposition 3.13]{All21} and compare also \cite{Raw22,Sch23}.

\section{ET-MHE Scheme}
\label{sec:MHEScheme} 
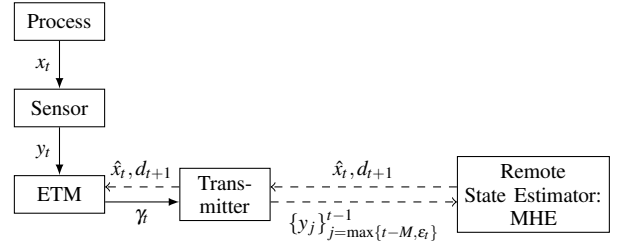
\begin{figure}[!t]
		\vspace{0.55em}
	\centering
	\setlength\figurewidth{0.9\columnwidth} 
\begin{tikzpicture}
	\usetikzlibrary{arrows}
	\usetikzlibrary{shapes.geometric}
\begin{footnotesize}
	\tikzstyle{block} = [draw, minimum width=1.2cm, minimum height=0.5cm,  
	align = center]
	[shorten -2pt,shorten -2pt,-implies]
	\node[block] (sys)  at (-3,3.3) {Process};
	\node[block] (sens)  at (-3,2.15) {Sensor};
	\node[block] (ES) at (-3,1) {ETM};
	\node[block] (TM) at (-0.8,1) {Trans-\\mitter};
	\node[block] (RO)  at (3.3,1) {Remote\\ State Estimator: \\ MHE};
	\draw[-latex] (sys) --(sens) node [left,midway] {$x_{t}$ };
	\draw[-latex] (sens) --(ES) node [left,midway] { $y_{t}$ };
	\draw[-latex] ([yshift=-1mm]ES.east) --([yshift=-1mm]TM.west) node [below,midway] { $\gamma_{t}$};
	\draw[dashed, ->] ([yshift=1mm]TM.west) --([yshift=1mm]ES.east) node [above,midway] {$\hat{x}_{t},d_{t+1}$ };
	\draw[dashed, ->] ([yshift=-1mm]TM.east) --([yshift=-1mm]RO.west) node [below,midway] {$\{y_j\}_{j=\max\{t-M,\epsilon_t\}}^{t-1}$};
	\draw[dashed, ->] ([yshift=1mm]RO.west) --([yshift=1mm]TM.east) node [above,midway] {$\hat{x}_{t}, d_{t+1}$};
\end{footnotesize}
\end{tikzpicture}
	\caption{Diagram of the overall ET-MHE framework. The dashed arrows represent the communication that is only required in case of an event.} 
	\label{fig:system}
	\vspace{-0.3em}
\end{figure}

MHE is an optimization-based state estimation technique that 
solves an optimization problem over a window of past inputs and outputs at each time $t$. 
In the following, we propose an ET-MHE scheme that requires to explicitly solve the optimization problem only at the time instances when an event occurs. In addition, such an ET-MHE reduces the frequency with which  a communication channel between the plant and a remote estimator needs to be accessed.  A diagram of the overall  ET-MHE framework is presented in Figure~\ref{fig:system}. 
\par The ETM determines the value of the binary event-triggering variable $\gamma_{t}$. If $\gamma_{t}=1$, an event is scheduled; $\gamma_t=0$ indicates that there is no event and thus no data transmission at time $t$.
The last time an event has occurred before the current time $t$ is denoted by 
 $\epsilon_t=\max\{0\leq\tau<t|\gamma_{\tau}=1\}$. For simplicity, we set $\epsilon_0=0$ and $\gamma_0=1$. Furthermore, let $\delta_t$  refer to the number of time steps  that have passed since the last event, i.e.,
\begin{align}
	\delta_{t}=	\begin{cases}
	t-\epsilon_t, & \gamma_t=0\\
	0, &\gamma_t=1.
\end{cases}
\label{eq:delta}
	\end{align}
	Note that if an event is scheduled at the current time $t$, then $\delta_t=0$. 	If $\gamma_{t}=1$, then the output sequence $\{y_j\}_{j=\max\{t-M,\epsilon_t\}}^{t-1}$ is sent
	to a remote estimator, where $M\in \mathbb{I}_{\geq 0}$  defines the horizon length of the MHE as described below. 
 How the ETM schedules an event will be discussed in more detail later in this section.
\par \textcolor{new}{The remote estimator consists of an MHE with a window length given by $M_t=\min\{t,M+\delta_t\}$ with $M\in \mathbb{I}_{\geq 0}$.
\begin{rem}
Note that in our scheme the MHE optimization  problem needs to be explicitly solved only when an event is triggered (see Proposition~\ref{prop:NLPOL} below). 
Since $\delta_t = 0$ in case of an event, it follows that we always explicitly solve the optimization problem with a fixed horizon length of $M$ for $t \geq M$. 
\label{rem:M}
\end{rem}}
 The MHE's nonlinear program (NLP) is given by 
\begin{subequations}
	\begin{align}
		\min_{\hat{x}_{t-M_t|t}, \hat{w}_{\cdot|t}} 
		&	J(\hat{x}_{t-M_t|t}, \hat{w}_{\cdot|t},\hat{y}_{\cdot|t},t)\\
		\text{s.t.} \  \hat{x}_{j+1|t}&=f(\hat{x}_{j|t},u_j,\hat{w}_{j|t}), \ j\in \mathbb{I}_{[t-M_t,t-1]},\label{eq:NLPC1}\\
		\hat{y}_{j |t}&=h(\hat{x}_{j|t},u_j,\hat{w}_{j|t}), \ j\in \mathbb{I}_{[t-M_t,t-1]},\label{eq:NLPC2}\\
		\hat{w}_{j|t} &\in \mathbb{W}, \hat{y}_{j|t} \in \mathbb{Y}, \ j\in \mathbb{I}_{[t-M_t,t-1]},\label{eq:NLPC3}\\
			\hat{x}_{j|t} &\in \mathbb{X}, \ j\in \mathbb{I}_{[t-M_t,t]}.\label{eq:NLPC4} 
	\end{align}
	\label{eq:ET2ETNLP}
\end{subequations}
The notation $\hat{x}_{j|t}$ denotes the estimated state for time $j$, computed at the (current) time $t$. The estimated disturbances and outputs are denoted analogously by $\hat{w}_{j|t}$ and $\hat{y}_{j|t}$. The optimal state, \mbox{disturbance} and output sequences minimizing the cost function  $J$ are denoted by $\hat{x}^*_{\cdot|t}$, $\hat{w}^*_{\cdot|t}$ and  $\hat{y}^*_{\cdot|t}$, respectively. We define the optimal estimate at the current time $t$  as $\hat{x}_t \coloneqq \hat{x}^*_{t|t}$. Moreover, $\hat{e}_t\coloneqq x_t-\hat{x}_t$ refers to the estimation error at time~$t$. 
We consider the following cost function
\begin{align}
	\begin{aligned}
			J(\hat{x}_{t-M_t|t}, \hat{w}_{\cdot|t},  \hat{y}_{\cdot|t},t)&=2\eta^{M_t}||\hat{x}_{t-M_t|t}-\hat{x}_{t-M_t}||_{P_2}^2 \\&+(\alpha+1)\Big(\sum_{j=t-M_t}^{t-1} \eta^{t-j-1}2 ||\hat{w}_{j|t}||_Q^2\\& +\sum_{j=t-M_t}^{t-\delta_t -1} \eta^{t-j-1} ||\hat{y}_{j|t}-y_j||_R^2 \Big).
	\end{aligned}
	\label{eq:objfunc}
\end{align}
The parameter $\alpha\geq0$ is a design variable and corresponds to the sensitivity of the event-triggering condition, see below for further discussion.
The cost function consists of two parts. The first term penalizes the difference between the first element of the estimated state sequence  $\hat{x}_{t-M_t|t}$ and the prior estimate $\hat{x}_{t-M_t}$ that was obtained  at time  $t-M_t$. This is the so called filtering prior (cf. \cite{Raw22}). The second part, the stage cost, penalizes the estimated noise and the difference between the measured and the estimated output. The weighting matrices $P_2,Q,$ and $R$ correspond to the matrices in Assumption \ref{ass:eIOSSET}. If the system is exponentially \iIOSS/, then the cost function can be arbitrarily parameterized with any positive definite matrices $P_2,Q$ and $R$  (since (\ref{eq:Lyap}) can be rescaled accordingly), however, the choice of $P_2$ influences the minimum horizon length required for the estimator to be RGES (cf. \cite[Remark 1]{Sch23}). 
Due to the discount factor $\eta$, disturbances and outputs that are more distant in the past have less influence on the cost function. An MHE cost function with such a discount factor was introduced in \cite{Knu18} and has proven to be very useful in the robust stability analysis of MHE. 
 \par  In principle, the NLP (\ref{eq:ET2ETNLP})  could be solved at every time instance $t$ using the time-varying horizon length $M_t$. However, the following proposition shows that this is equivalent to explicitly solving (\ref{eq:ET2ETNLP})  only when an event is triggered, and using open-loop predictions when there is no event. 
\begin{prop}
	The solution of the NLP (\ref{eq:ET2ETNLP}) at time $t\geq0$ is given by 
	\begin{subequations}
	\begin{align}
		\hat{x}^*_{t-M_t|t} &= \hat{x}^*_{t-\delta_t-M_{t-\delta_t}|t-\delta_t},\\
		\hat{w}^*_{j|t} &= \hat{w}^*_{j|t-\delta_t}, \ j\in [t-M_t,t-\delta_t-1],\\
		\hat{w}^*_{j|t} &= 0, \ j \in [t-\delta_t,t-1]. \label{eq:Propw0}
	\end{align}
	\label{eq:Prop}
	\end{subequations}
\label{prop:NLPOL}	
\end{prop}
\begin{proof}
First,	recall that the last term of the cost function only considers the time instances when the respective measurements were sent to the remote estimator, i.e, the measurements in the interval $[t-M_t,t-\delta_t-1]$. Hence,   $\hat{w}_{j|t}=0$ for all \mbox{$j\in [t-\delta_t,t-1]$} minimizes $J(\hat{x}_{t-M_t|t}, \hat{w}_{\cdot|t},  \hat{y}_{\cdot|t},t)$. 
Therefore, the optimal value of the cost function at time $t$ satisfies
\begin{align*}
	\begin{aligned}
		&J(\hat{x}_{t-M_t|t}^*, \hat{w}_{\cdot|t}^*,  \hat{y}_{\cdot|t}^*,t)\\=&\eta^{\delta_t} J(\hat{x}_{t-\delta_t-M_{t-\delta_t}|t-\delta_t}^*, \hat{w}_{\cdot|t-\delta_t}^*,  \hat{y}_{\cdot|t-\delta_t}^*,t-\delta_t)
	\end{aligned}
\end{align*}  
where  $t-\delta_t-M_{t-\delta_t}=t-M_t$ since $\delta_{t-\delta_t} =0$. Thus, the solution of minimizing  $J(\hat{x}_{t-M_t|t}, \hat{w}_{\cdot|t},\hat{y}_{\cdot|t},t)$ is given by (\ref{eq:Prop}). 
	\end{proof}
	\par Proposition~\ref{prop:NLPOL} implies that if $\gamma_t=0$, i.e., no event is scheduled, 
	the estimate $\hat{x}_j$ is given for all $j\in [t-\delta_t+1,t]$ by 
	\begin{align}
		\hat{x}_{j}=f(\hat{x}_{j-1},u_{j-1},0),
		\label{eq:OL}
	\end{align}
	which corresponds to an open-loop prediction. Therefore, it suffices to explicitly solve the  NLP~(\ref{eq:ET2ETNLP}) only if an event is triggered. Recall that if $\gamma_t=1$, then $\delta_t=0$. Hence, \textcolor{new}{as discussed} in Remark \ref{rem:M}, the NLP that needs to be explicitly solved has a fixed horizon length $M$ for all $t\geq M$.
\par The ETM determines the value of the scheduling variable $\gamma_t$ at each time as follows
\begin{align}
		\gamma_{t}=	\begin{cases}
			0, & \text{if} \  \begin{aligned}&\sum_{j=\epsilon_t}^{t-1}\eta^{t-j-1} ||y_j-h(\hat{x}_{j},u_j,0)||_R^2 \\&< \alpha \eta^{t-\epsilon_t}d_t
			\end{aligned}\\
			1, &\text{otherwise}
		\end{cases}
		\label{eq:ET2trigruled}
	\end{align}
	with  $\alpha \in \mathbb{R}_{\geq 0}$ and
		\begin{align}
		\begin{aligned}
			d_{t}=	\sum_{j=\epsilon_t-M_{\epsilon_{t}}}^{\epsilon_t-1} \eta^{\epsilon_t-1-j}\Big(2 ||\hat{w}_{j|\epsilon_t}^*||_Q^2+ ||\hat{y}_{j|\epsilon_t}^*-y_j||_R^2\Big),
		\end{aligned}
		\label{eq:dwy}
	\end{align}
where $M_{\epsilon_t}$ is the horizon length $M_t$ for $t=\epsilon_t$. As previously stated, if $\gamma_{t}=1$, the measurement sequence $\{y_j\}_{j=\max\{t-M,\epsilon_t\}}^{t-1}$ is transmitted to the remote state estimator and the optimization problem is solved at time $t$. 
Then, $d_{t+1}$ can be computed based on the estimated disturbance and output sequences $\hat{w}_{\cdot|t}^*, \hat{y}_{\cdot|t}^*$ since $\epsilon_{t+1}=t$.  Both $d_{t+1}$  and $\hat{x}_t$ are sent back to the ETM to evaluate the triggering condition for the next time step (cf. Figure \ref{fig:system}).
Note that $d_t$ is constant between events, i.e., $d_{t+1}=d_t$ if $\gamma_t=0$. This follows from (\ref{eq:Prop}) and  the fact that $\epsilon_{t+1}=\epsilon_t$ if $\gamma_t=0$. The  steps of the ET-MHE scheme are outlined in  Algorithm \ref{alg}. 
\floatstyle{spaceruled}
\restylefloat{algorithm}
	\begin{algorithm}[!t]
	\caption{Event-triggered MHE}
	\begin{algorithmic}[1]
		\State Set $\gamma_0=1$ and $d_{1}=0$.
		\State Set $t=1$.
		\State ETM computes $\gamma_{t}$.
		\If{$\gamma_{t}=1$}
		\State $\{y_j\}_{j=\max\{t-M,\epsilon_t\}}^{t-1}$ is sent to remote estimator.
		\State NLP (\ref{eq:ET2ETNLP}) of MHE is solved.
		\State $d_{t+1}$ is calculated.
		\State $d_{t+1}$ and $\hat{x}_t$ are sent back to the ETM. 
		\Else 
		\State   $\hat{x}_t$ is calculated according to (\ref{eq:OL}). 
		\EndIf
		\State  $\hat{y}_t=h(\hat{x}_t,u_t,0)$ is calculated.
		\State Set $t=t+1$ and go back to step 3.
	\end{algorithmic}
	\label{alg}
\end{algorithm}
 \begin{rem}
 	In the proposed method, $\alpha$ is  a design parameter that influences how frequently an event is scheduled. The larger $\alpha$, the fewer events will occur and the larger the disturbance gain of the error bound (see Theorem \ref{thm:stab}).
 \end{rem}
\section{Stability Analysis}
	\label{sec:StabAna}
		Having introduced the ET-MHE scheme, we now prove robust stability of the proposed scheme. The proof of the following theorem is based on the approach in \cite {Sch23} combined with the event-triggering condition  (\ref{eq:ET2trigruled}). Additionally, adaptations are necessary due to the modified formulation of the cost function (\ref{eq:objfunc}) regarding the horizon length. 
		\begin{thm}[RGES of ET-MHE]
			Let Assumption \ref{ass:eIOSSET} hold and the horizon $M\in \mathbb{I}_{\geq 0}$ be chosen such that $4\lambda_{\text{max}}(P_2,P_1)\eta^{M}<1$. Then there exists $\rho \in [0,1)$ such that the state estimation error of the ET-MHE
			scheme~(\ref{eq:ET2ETNLP}) with the event scheduling condition (\ref{eq:ET2trigruled}) satisfies for all $t\geq 0$ 
			\begin{align*}
				\begin{aligned}
						||\hat{e}_t||&\leq	2 \sqrt{\frac{\lambda_{\text{max}}(P_2)}{\lambda_{\text{min}}(P_1)}}\sqrt{\rho}^{t}||\hat{e}_{0}|| \\&+\sqrt{\frac{(2\alpha+4)\lambda_{\text{max}}(Q)}{\lambda_{\text{min}}(P_1)}}\sum_{q=0}^{t-1} \sqrt{\rho}^{t-q-1}  ||w_q||,
				\end{aligned}
			\end{align*}
				i.e., the ET-MHE is an RGES estimator according to Definition \ref{def:rges}.
			\label{thm:stab}
		\end{thm}
		
		\begin{proof}
			Due to the constraints (\ref{eq:NLPC1})-(\ref{eq:NLPC4}) in the NLP (\ref{eq:ET2ETNLP}), at each time step $t$, the estimated trajectories satisfy (\ref{eq:sys}), $\hat{x}_{j|t}\in \mathbb{X}$ for all $j\in \mathbb{I}_{[t-M_t,t]}$ and $\hat{w}_{j|t}\in \mathbb{W},\hat{y}_{j|t}\in \mathbb{Y}$ for all $j\in \mathbb{I}_{[t-M_t,t-1]}$. Thus, we can apply (\ref{eq:ioss}) to obtain
			\begin{align*}
				\begin{aligned}
					||\hat{x}_t-x_t||_{P_1}^2 &\leq \eta^{M_t} ||\hat{x}_{t-M_t|t}^*-x_{t-M_t}||_{P_2}^2\\&+\sum_{j=t-M_t}^{t-1}\eta^{t-j-1}||\hat{w}_{j|t}^*-w_j||_Q^2\\&
					 +\sum_{j=t-M_t}^{t-1}\eta^{t-j-1}||\hat{y}_{j|t}^*-y_j||_R^2. 
				\end{aligned}
			\end{align*}
			Applying the Cauchy-Schwarz inequality, and Young’s inequality, it holds that
			\begin{align*}
				||\hat{w}_{j|t}^*-w_j||_Q^2\leq 2||\hat{w}_{j|t}^*||_Q^2+ 2||w_j||_Q^2
			\end{align*}
		and therefore,
					\begin{align}
					\begin{aligned}
						||\hat{x}_t-x_t||_{P_1}^2 &\leq 
						\eta^{M_t} ||\hat{x}_{t-M_t|t}^*-x_{t-M_t}||_{P_2}^2\\&+\sum_{j=t-M_t}^{t-1}\eta^{t-j-1} 2||w_j||_Q^2\\
						&+\sum_{j=t-M_t}^{t-1}\eta^{t-j-1}2||\hat{w}_{j|t}^*||_Q^2\\&	 +\sum_{j=t-M_t}^{t-1}\eta^{t-j-1}||\hat{y}_{j|t}^*-y_j||_R^2.
					\end{aligned}
					\label{eq:ET2eiossWsplit}
				\end{align}
				\par Recall that the latest event in  the time interval $[0,t]$ was scheduled at time \mbox{$t-\delta_t$}. Thus, the output sequence  $\{y_j\}_{j=t-M_t}^{t-\delta_t-1}$ was sent to the remote estimator. 
				From Proposition~\ref{prop:NLPOL} it follows that $\hat{y}_{j|t}^*=h(\hat{x}_{j},u_j,0)$ for all $j \in [t-\delta_t,t-1]$ and that
				\begin{align*}
					\sum_{j=t-M_t}^{t-1}\eta^{t-j-1}2||\hat{w}_{j|t}^*||_Q^2=\sum_{j=t-M_t}^{t-\delta_t-1}\eta^{t-j-1}2||\hat{w}_{j|t-\delta_t}^*||_Q^2.
				\end{align*}
				Hence, we can write 
					\begin{align}
					\begin{aligned}
						&||\hat{x}_t-x_t||_{P_1}^2 \leq 
						\eta^{M_t} ||\hat{x}_{t-M_t|t}^*-x_{t-M_t}||_{P_2}^2\\&+\sum_{j=t-M_t}^{t-1}\eta^{t-j-1} 2||w_j||_Q^2
						\\&+\sum_{j=t-M_t}^{t-\delta_t-1}\eta^{t-j-1}2||\hat{w}_{j|t-\delta_t}^*||_Q^2\\&	+\sum_{j=t-M_t}^{t-\delta_t-1}\eta^{t-j-1}||\hat{y}_{j|t-\delta_t}^*-y_j||_R^2\\&+\sum_{j=t-\delta_t}^{t-1}\eta^{t-j-1}||y_j-h(\hat{x}_{j},u_j,0)||_R^2.
					\end{aligned}
					\label{eq:ET2eiossEstsplit}
				\end{align} 
			\par If $\gamma_{t}=1$, then the measurement sequence $\{y_j\}_{j=\max\{t-M,\epsilon_t\}}^{t-1}$ is sent to the MHE and the optimization problem is solved at time $t$. Hence, the estimation error is  bounded by (\ref{eq:ET2eiossEstsplit}) with $\delta_t=0$, resulting in (\ref{eq:ET2eiossWsplit}) again.
			\par Recall that if no event is triggered at time $t$, then $\epsilon_t=t-\delta_ t$ (cf. (\ref{eq:delta})) and thus $\epsilon_t-M_{\epsilon_t}=t-M_t$. Hence, if $\gamma_t=0$, i.e., according to (\ref{eq:ET2trigruled}) and (\ref{eq:dwy}) 
			\begin{align*}
				\begin{aligned}
					&\sum_{j=t-\delta_t}^{t-1}\eta^{t-j-1} ||y_j-h(\hat{x}_j,u_j,0)||_R^2 \\&< \alpha\sum_{j=t-M_t}^{t-\delta_t-1} \eta^{t-j-1} \Big( 2 ||\hat{w}_{j|t-\delta_t}^*||_Q^2+||\hat{y}_{j|t-\delta_t}^*-y_j||_R^2\Big)
				\end{aligned}
			\end{align*}
			then, the bound in  (\ref{eq:ET2eiossEstsplit}) can be upper bounded as follows
			\begin{align*}
				\begin{aligned}
					||\hat{x}_t-x_t||_{P_1}^2 &\leq \eta^{M_t} ||\hat{x}_{t-M_t|t}^*-x_{t-M_t}||_{P_2}^2\\&+\sum_{j=t-M_t}^{t-1}\eta^{t-j-1} 2||w_j||_Q^2\\
					&+(\alpha+1)\sum_{j=t-M_t}^{t-\delta_t-1}\eta^{t-j-1} 2 ||\hat{w}_{j|t-\delta_t}^*||_Q^2\\ 
				&+(\alpha+1)\sum_{j=t-M_t}^{t-\delta_t-1}\eta^{t-j-1}||\hat{y}_{j|t-\delta_t}^*-y_j||_R^2.
				\end{aligned}
			\end{align*}
			\par Due to 	\begin{align*}
				\begin{aligned}
					&||\hat{x}_{t-M_t|t}^*-x_{t-M_t}||_{P_2}^2\\=& ||\hat{x}_{t-M_t}-x_{t-M_t}+\hat{x}_{t-M_t|t}^*-\hat{x}_{t-M_t}||_{P_2}^2\\
					\leq &2||\hat{x}_{t-M_t}-x_{t-M_t}||_{P_2}^2+2||\hat{x}_{t-M_t|t}^*-\hat{x}_{t-M_t}||_{P_2}^2
				\end{aligned}
			\end{align*}
			we then obtain
			\begin{align*}
				\begin{aligned}
					||\hat{x}_t-x_t||_{P_1}^2 &\leq 2 \eta^{M_t} ||\hat{x}_{t-M_t}-x_{t-M_t}||_{P_2}^2\\&+\sum_{j=t-M_t}^{t-1}2\eta^{t-j-1} ||w_j||_Q^2 +J(\hat{x}_{t-M_t|t}^*, \hat{w}_{\cdot|t}^*,  \hat{y}_{\cdot|t}^*,t) .
				\end{aligned}
			\end{align*}
			Since $J(\hat{x}_{t-M_t|t}^*, \hat{w}_{\cdot|t}^*, \hat{y}_{\cdot|t}^*,t) \leq J(x_{t-M_t}, w_{\cdot|t}, y_{\cdot|t},t)$	by optimality, it holds that
			\begin{align}
				\begin{aligned}
					||\hat{x}_t-x_t||_{P_1}^2 &\leq 4\eta^{M_t} ||\hat{x}_{t-M_t}-x_{t-M_t}||_{P_2}^2\\&+(2\alpha+4)\sum_{j=t-M_t}^{t-1}\eta^{t-j-1} ||w_j||_Q^2.
				\end{aligned}
				\label{eq:ET2boundMtstep1ET}
			\end{align} 
		Due to $||\hat{x}_{t-M_t}-x_{t-M_t}||_{P_2}^2\leq\lambda_{\text{max}}(P_2,P_1)||\hat{x}_{t-M_t}-x_{t-M_t}||_{P_1}^2$ we can write
			\begin{align*}
				\begin{aligned}
					||\hat{x}_t-x_t||_{P_1}^2 
					&\leq 4 \eta^{M_t} \lambda_{\text{max}}(P_2,P_1)||\hat{x}_{t-M_t}-x_{t-M_t}||_{P_1}^2\\&+(2\alpha+4)\sum_{j=t-M_t}^{t-1}\eta^{t-j-1} ||w_j||_Q^2.
				\end{aligned}
			\end{align*}
			\textcolor{new}{We select the horizon length  $M$ large enough such that 
			\begin{align}
				\rho^M:=4\lambda_{\text{max}}(P_2,P_1)\eta^{M}<1
				\label{eq:minhor}
			\end{align}
			with $\rho \in [0,1)$. Noting that since $\lambda_{\text{max}}(P_2,P_1)\geq 1$, it holds  that $\rho\geq\eta$}, and recalling that $M_t \geq M$ for all $t\geq M$, we obtain for all $t\geq M$
			\begin{align}
				\begin{aligned}
					||\hat{e}_t||_{P_1}^2=	||\hat{x}_t-x_t||_{P_1}^2 &\leq \rho^{M_t} ||\hat{x}_{t-M_t}-x_{t-M_t}||_{P_1}^2\\&+(2\alpha+4)\sum_{j=t-M_t}^{t-1}\eta^{t-j-1} ||w_j||_Q^2. 
				\end{aligned}
				\label{eq:ET2boundMstepET}
			\end{align}
			Consider some time $t=l+T$, with 	$l\in \mathbb{I}_{[0,M-1]}$ and \mbox{$T\in \mathbb{I}_{\geq M}$} consisting of $k$  time intervals $[t_{i+1},t_i]$, $i=1,\ldots, k$ with \mbox{$t_{i+1}=t_i-M-\delta_{t_i}$}, $t_{1}=t$ and $t_{k+1}=l$. 
			Using (\ref{eq:ET2boundMtstep1ET}) we obtain
			\begin{align}
				\begin{aligned}
					||\hat{e}_l||_{P_1}^2 &\leq  4 \eta^{l} ||\hat{e}_{0}||_{P_2}^2+(2\alpha+4)\sum_{j=0}^{l-1}\eta^{l-j-1} ||w_j||_Q^2.
				\end{aligned}
				\label{eq:ET2boundlkET}
			\end{align}
			Applying (\ref{eq:ET2boundMstepET})  for each of the above $k$ time intervals yields
			\begin{align*}
				\begin{aligned}
					||\hat{e}_{t_i}||_{P_1}^2&\leq \rho^{M+\delta_{t_i}} ||\hat{e}_{t_{i+1}}||_{P_1}^2\\&+(2\alpha+4)\sum_{j=t_i-M-\delta_{t_i}}^{t_i-1}\eta^{t_i-j-1} ||w_j||_Q^2.
				\end{aligned}
			\end{align*}
			Hence, we obtain the following upper bound on the estimation error at time $t$
			\begin{align*}
				\begin{aligned}
					||\hat{e}_t||_{P_1}^2 &\leq \rho^{T}	||\hat{e}_l||_{P_1}^2+(2\alpha+4)\sum_{j=l}^{t-1}\eta^{t-j-1}||w_j||_Q^2 \\
					&\stackrel{(\ref{eq:ET2boundlkET})}{\leq} \rho^{T}\Big(4\eta^{l} ||\hat{e}_{0}||_{P_2}^2+(2\alpha+4)\sum_{j=0}^{l-1}\eta^{l-j-1} ||w_j||_Q^2 \Big) \\&+(2\alpha+4)\sum_{j=l}^{t-1}\eta^{t-j-1}||w_j||_Q^2 .
				\end{aligned}
			\end{align*}
			Due to $\eta\leq \rho$ we can write 
			
			\begin{align*}
				\begin{aligned}
					||\hat{e}_t||_{P_1}^2&\leq  4 \rho^{t}||\hat{e}_{0}||_{P_2}^2+(2\alpha+4)\sum_{q=0}^{t-1} \rho^{t-q-1}  ||w_q||_Q^2.
				\end{aligned}
			\end{align*}
			Since $\lambda_{\text{min}}(P_1)||\hat{e}_t||^2\leq ||\hat{e}_t||_{P_1}^2$, $||\hat{e}_0||_{P_2}^2\leq \lambda_{\text{max}}(P_2)||\hat{e}_0||^2$, and   $||\hat{w}_q||_{Q}^2\leq \lambda_{\text{max}}(Q)||\hat{w}_q||^2$ it holds that
			\begin{align*}
				\begin{aligned}
					||\hat{e}_t||^2&\leq  4 \frac{\lambda_{\text{max}}(P_2)}{\lambda_{\text{min}}(P_1)} \rho^{t}||\hat{e}_{0}||^2\\&+\frac{(2\alpha+4)\lambda_{\text{max}}(Q)}{\lambda_{\text{min}}(P_1)}\sum_{q=0}^{t-1} \rho^{t-q-1}  ||w_q||^2.
				\end{aligned}
			\end{align*}	
			Using $\sqrt{a+b}\leq \sqrt{a}+\sqrt{b}$ for all $a,b\geq0$ results in 
			\begin{align*}
				\begin{aligned}
					||\hat{e}_t||&\leq	2 \sqrt{\frac{\lambda_{\text{max}}(P_2)}{\lambda_{\text{min}}(P_1)}}\sqrt{\rho}^{t}||\hat{e}_{0}|| \\&+\sqrt{\frac{(2\alpha+4)\lambda_{\text{max}}(Q)}{\lambda_{\text{min}}(P_1)}}\sum_{q=0}^{t-1} \sqrt{\rho}^{t-q-1}  ||w_q||
				\end{aligned}
			\end{align*}
			and concludes the proof.	
		\end{proof}
		\section{Numerical Example}
		\label{sec:NumEx}
		In this section, we illustrate our results by applying the proposed method to the following system
		\begin{align*}
			\begin{aligned}
			x_{1,t+1}&=x_1+\tau(-2k_1x_{1,t}^2+2k_2x_{2,t})+w_{1,t},\\
			x_{2,t+1}&=x_2+\tau(k_1x_{1,t}^2-k_2x_{2,t})+w_{2,t},\\
			y&=x_{1,t}+x_{2,t}+w_{3,t}
				\end{aligned}
		\end{align*}
		with $k_1=0.16,\ k_2=0.0064$, and sampling time $\tau=0.1$. This corresponds to a batch reactor system from \cite{Ten02} using an Euler discretization. This example is frequently used in the MHE literature as a benchmark (e.g. in \cite{Raw22,Sch23}) since other nonlinear state estimators, such as the extended Kalman filter, can fail to provide meaningful results. We consider \mbox{$x_0=[3,1]^\top$} and the poor initial estimate $\hat{x}_0=[0.1,4.5]^\top$. In the simulation, the  additional disturbance $w\in \mathbb{R}^3$ is treated as  a uniformly distributed random variable that satisfies $|w_i|\leq 10^{-3}, \ i=1,2$ and $|w_3| \leq 0.1$. 
		\par
\begin{figure}[!t]
					\vspace{0.6em}
				\centering
				\centerline{\input{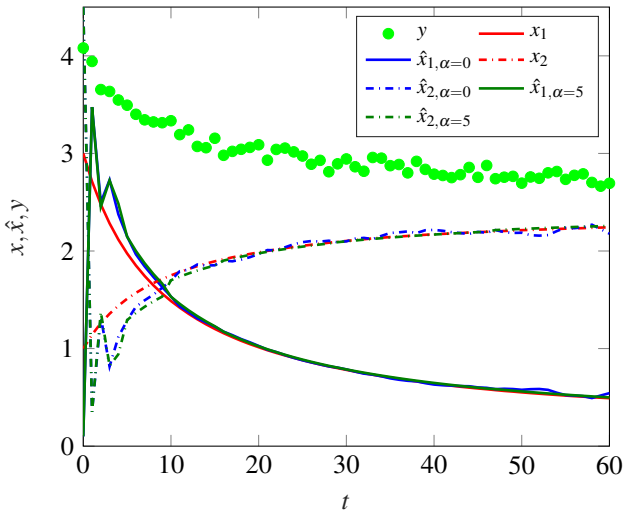}}
				\caption{Comparison of ET-MHE results for $\alpha=5$, MHE estimates (without event-triggering, i.e., $\alpha=0$) and real system states.} 
				\label{fig:simmhe}
			\end{figure}

\begin{figure}[!t]
				\centering
				\centerline{
%
%
\definecolor{mycolor1}{rgb}{0.00000,0.44700,0.74100}%
\definecolor{mycolor2}{rgb}{0.85000,0.32500,0.09800}%
\definecolor{mycolor3}{rgb}{0.92900,0.69400,0.12500}%
\definecolor{mycolor4}{rgb}{0.49400,0.18400,0.55600}%
\definecolor{mycolor5}{rgb}{0.46600,0.67400,0.18800}%
\definecolor{mycolor6}{rgb}{0.30100,0.74500,0.93300}%
\definecolor{mycolor7}{rgb}{0.63500,0.07800,0.18400}%
\begin{tikzpicture}

\begin{axis}[%
width=0.9\figurewidth,
height=1.76in,
scale only axis,
xmin=0,
xmax=60,
xlabel style={font=\color{white!15!black}},
xlabel={$t$},
ymin=0.1,
ymax=10,
ylabel style={font=\color{white!15!black},at={(axis description cs:0.05,0.5)}},
ytick={1,3,5,7,9},
yticklabels={$\alpha=0.1$, $\alpha=1$,  $\alpha=3$, $\alpha=5$, $\alpha=20$,},
axis background/.style={fill=white}
]
\addplot [color=mycolor1, line width=0.1pt, only marks, mark=*, mark options={solid, mycolor1}, forget plot]
  table[row sep=crcr]{%
1	1\\
2	1\\
3	1\\
4	1\\
5	1\\
6	1\\
7	0\\
8	0\\
9	1\\
10	1\\
11	1\\
12	1\\
13	1\\
14	1\\
15	1\\
16	1\\
17	1\\
18	0\\
19	0\\
20	1\\
21	1\\
22	1\\
23	1\\
24	1\\
25	0\\
26	0\\
27	1\\
28	0\\
29	1\\
30	0\\
31	1\\
32	0\\
33	1\\
34	1\\
35	1\\
36	0\\
37	0\\
38	1\\
39	1\\
40	0\\
41	1\\
42	1\\
43	1\\
44	0\\
45	1\\
46	1\\
47	1\\
48	1\\
49	0\\
50	0\\
51	1\\
52	0\\
53	0\\
54	1\\
55	1\\
56	0\\
57	0\\
58	1\\
59	1\\
60	1\\
61	0\\
62	1\\
63	0\\
64	0\\
65	1\\
66	0\\
67	0\\
68	1\\
69	1\\
70	0\\
71	1\\
72	0\\
73	1\\
74	0\\
75	0\\
76	1\\
77	1\\
78	1\\
79	0\\
80	1\\
81	0\\
82	1\\
83	1\\
84	1\\
85	1\\
86	1\\
87	1\\
88	0\\
89	0\\
90	0\\
91	1\\
92	1\\
93	1\\
94	0\\
95	0\\
96	1\\
97	0\\
98	1\\
99	1\\
100	1\\
};
\addplot [color=mycolor1, line width=0.1pt, only marks, mark=*, mark options={solid, mycolor1}, forget plot]
  table[row sep=crcr]{%
1	3\\
2	3\\
3	3\\
4	3\\
5	0\\
6	3\\
7	0\\
8	0\\
9	0\\
10	3\\
11	3\\
12	0\\
13	0\\
14	3\\
15	0\\
16	0\\
17	3\\
18	0\\
19	0\\
20	0\\
21	3\\
22	0\\
23	0\\
24	0\\
25	3\\
26	0\\
27	0\\
28	0\\
29	3\\
30	0\\
31	0\\
32	0\\
33	0\\
34	0\\
35	3\\
36	0\\
37	0\\
38	0\\
39	0\\
40	0\\
41	0\\
42	3\\
43	0\\
44	0\\
45	0\\
46	0\\
47	3\\
48	0\\
49	0\\
50	0\\
51	0\\
52	3\\
53	0\\
54	0\\
55	0\\
56	0\\
57	3\\
58	0\\
59	0\\
60	3\\
61	0\\
62	0\\
63	0\\
64	0\\
65	3\\
66	0\\
67	0\\
68	0\\
69	0\\
70	3\\
71	0\\
72	0\\
73	0\\
74	0\\
75	0\\
76	3\\
77	3\\
78	0\\
79	0\\
80	0\\
81	0\\
82	0\\
83	3\\
84	0\\
85	0\\
86	0\\
87	0\\
88	0\\
89	0\\
90	0\\
91	0\\
92	3\\
93	0\\
94	0\\
95	0\\
96	3\\
97	0\\
98	0\\
99	0\\
100	3\\
};
\addplot [color=mycolor1, line width=0.1pt, only marks, mark=*, mark options={solid, mycolor1}, forget plot]
  table[row sep=crcr]{%
1	5\\
2	5\\
3	5\\
4	0\\
5	5\\
6	0\\
7	0\\
8	0\\
9	0\\
10	5\\
11	0\\
12	0\\
13	5\\
14	0\\
15	0\\
16	0\\
17	5\\
18	0\\
19	0\\
20	0\\
21	0\\
22	0\\
23	0\\
24	5\\
25	0\\
26	0\\
27	0\\
28	0\\
29	0\\
30	0\\
31	0\\
32	0\\
33	0\\
34	0\\
35	5\\
36	0\\
37	0\\
38	0\\
39	0\\
40	0\\
41	0\\
42	0\\
43	0\\
44	0\\
45	0\\
46	0\\
47	5\\
48	0\\
49	0\\
50	0\\
51	0\\
52	0\\
53	0\\
54	0\\
55	0\\
56	0\\
57	0\\
58	0\\
59	0\\
60	5\\
61	0\\
62	0\\
63	0\\
64	0\\
65	0\\
66	0\\
67	0\\
68	0\\
69	5\\
70	0\\
71	0\\
72	0\\
73	0\\
74	0\\
75	0\\
76	0\\
77	0\\
78	0\\
79	5\\
80	0\\
81	0\\
82	0\\
83	0\\
84	0\\
85	5\\
86	0\\
87	0\\
88	0\\
89	0\\
90	0\\
91	0\\
92	0\\
93	0\\
94	0\\
95	0\\
96	0\\
97	0\\
98	0\\
99	0\\
100	5\\
};
\addplot [color=mycolor1, line width=0.1pt, only marks, mark=*, mark options={solid, mycolor1}, forget plot]
  table[row sep=crcr]{%
1	7\\
2	7\\
3	7\\
4	0\\
5	7\\
6	0\\
7	0\\
8	0\\
9	0\\
10	7\\
11	0\\
12	0\\
13	0\\
14	0\\
15	0\\
16	7\\
17	0\\
18	0\\
19	0\\
20	0\\
21	0\\
22	0\\
23	0\\
24	0\\
25	0\\
26	0\\
27	0\\
28	0\\
29	7\\
30	0\\
31	0\\
32	0\\
33	0\\
34	0\\
35	0\\
36	0\\
37	0\\
38	0\\
39	0\\
40	0\\
41	0\\
42	0\\
43	0\\
44	0\\
45	0\\
46	0\\
47	7\\
48	0\\
49	0\\
50	0\\
51	0\\
52	0\\
53	0\\
54	0\\
55	0\\
56	0\\
57	0\\
58	0\\
59	0\\
60	0\\
61	0\\
62	7\\
63	0\\
64	0\\
65	0\\
66	0\\
67	0\\
68	0\\
69	0\\
70	7\\
71	0\\
72	0\\
73	0\\
74	0\\
75	0\\
76	0\\
77	0\\
78	0\\
79	0\\
80	0\\
81	0\\
82	0\\
83	7\\
84	0\\
85	0\\
86	0\\
87	0\\
88	0\\
89	0\\
90	0\\
91	0\\
92	0\\
93	0\\
94	0\\
95	0\\
96	0\\
97	0\\
98	0\\
99	0\\
100	7\\
};
\addplot[color=mycolor1, line width=0.1pt, only marks, mark=*, mark options={solid, mycolor1}, forget plot]
  table[row sep=crcr]{%
1	9\\
2	9\\
3	9\\
4	0\\
5	0\\
6	0\\
7	9\\
8	0\\
9	0\\
10	0\\
11	0\\
12	0\\
13	0\\
14	0\\
15	0\\
16	0\\
17	0\\
18	0\\
19	0\\
20	9\\
21	0\\
22	0\\
23	0\\
24	0\\
25	0\\
26	0\\
27	0\\
28	0\\
29	0\\
30	0\\
31	0\\
32	0\\
33	0\\
34	0\\
35	0\\
36	0\\
37	0\\
38	0\\
39	0\\
40	0\\
41	0\\
42	0\\
43	0\\
44	0\\
45	0\\
46	0\\
47	9\\
48	0\\
49	0\\
50	0\\
51	0\\
52	0\\
53	0\\
54	0\\
55	0\\
56	0\\
57	0\\
58	0\\
59	0\\
60	0\\
61	0\\
62	0\\
63	0\\
64	0\\
65	0\\
66	0\\
67	0\\
68	0\\
69	0\\
70	0\\
71	0\\
72	0\\
73	9\\
74	0\\
75	0\\
76	0\\
77	0\\
78	0\\
79	0\\
80	0\\
81	0\\
82	0\\
83	0\\
84	0\\
85	0\\
86	0\\
87	0\\
88	0\\
89	0\\
90	0\\
91	0\\
92	0\\
93	9\\
94	0\\
95	0\\
96	0\\
97	0\\
98	0\\
99	0\\
100	0\\
};
\end{axis}

\end{tikzpicture}%
}
				\caption{Times an event was scheduled ($\gamma=1$) for different values of $\alpha$.} 
				\label{fig:gamma}
				\vspace{-0.3em}
			\end{figure}
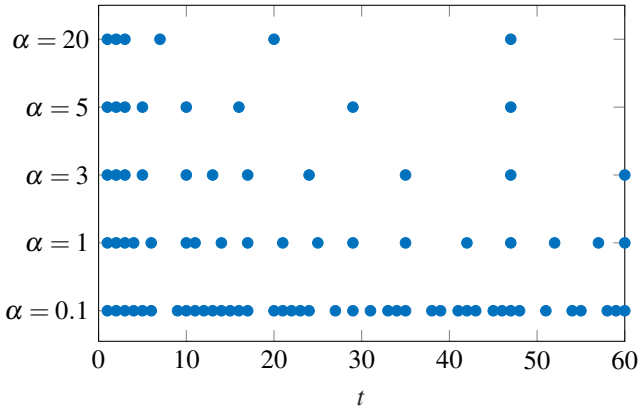

 For parameterizing the cost function we use the parameters of the quadratic  Lyapunov function $W_{\delta}=||x-\tilde{x}||_P^2$ computed in \cite{Sch23}  where 
		\begin{align*}
			P=\begin{bmatrix}4.539&4.171\\4.171&3.834
				\end{bmatrix},
				\ Q=\begin{bmatrix}10^3 & 0&0\\0& 10^4 &0\\0 & 0& 10^3 \end{bmatrix}, \ R=10^3,	
				\end{align*}
			$P_1=P_2=P$ and the decay rate $\eta=0.91$. 
			These parameters satisfy Assumption \ref{ass:eIOSSET}.
			Using condition (\ref{eq:minhor}) we obtain a minimal horizon length of  $M_{\text{min}}=15$ guaranteeing robust stability of the ET-MHE\textcolor{new}{\footnote{\textcolor{new}{Note that $\lambda_{\text{max}}(P,P)=1$ and thus $M_{\text{min}}>\log_{\eta}(\frac{1}{4})$.}}}. However, for the following simulation we select $M=30$ to increase the estimation performance. The simulation results for $\alpha=5$ are shown in Figure \ref{fig:simmhe}. 
			It is evident that the event-triggering does not significantly worsen the estimation performance.
			The choice of $\alpha$ influences how frequent an event is scheduled and thus the optimization problem needs to be solved. Figure~\ref{fig:gamma} depicts how the number  of events decreases with  increasing values of $\alpha$, which is in accordance with the ETM (\ref{eq:ET2trigruled}) (compare Remark 1).  For a fixed value of $\alpha$, increasing the horizon length tends to reduce the number of events, since $d_{t}$ as defined in (\ref{eq:dwy}) is, in general, larger in this case. For the selected parameterization $M=30$ and $\alpha=5$ the NLP needs to be solved only about 14\% of the time (cf. Figure~\ref{fig:gamma}).
		\section{Conclusion}
		\label{sec:con}
		This paper presents an event-triggered moving horizon state estimator. 
		The ETM and the MHE objective are designed such that both the optimization problem only needs to be solved and measurements only need to be sent to the remote estimator if an event is scheduled. When an event occurs a measurement sequence is sent to the estimator. Then,  the optimization problem is solved, and the current estimate is sent back to the plant side to be used in the ETM for scheduling the next event. Meanwhile, until the next event, state estimates are obtained by an open-loop prediction. 
	 	Furthermore, we  showed that under the assumption of the system being exponentially \iIOSS/ the proposed ET-MHE is robustly globally exponentially stable. The applicability of the ET-MHE was illustrated by a numerical example which in particular showed the potential of the method to significantly reduce the computational effort and the frequency of communication. In the presented example, these factors were reduced by approximately 86\%.

\end{document}